# WAKEFIELD DAMPING FOR THE CLIC CRAB CAVITY

P. K. Ambattu, G. Burt, A. C. Dexter and R. G. Carter, Cockcroft Institute, Lancaster University, Lancaster, UK, LA1 4YR

V. Khan and R.M. Jones, Cockcroft Institute, Manchester University, Warrington, UK, WA4 4AD

V. Dolgashev, SLAC, Menlo Park, CA 94025

*Abstract*

A crab cavity is required in the CLIC to allow effective head-on collision of bunches at the IP. A high operating frequency is preferred as the deflection voltage required for a given rotation angle and the RF phase tolerance for a crab cavity are inversely proportional to the operating frequency. The short bunch spacing of the CLIC scheme and the high sensitivity of the crab cavity to dipole kicks demand very high damping of the inter-bunch wakes, the major contributor to the luminosity loss of colliding bunches. This paper investigates the nature of the wakefields in the CLIC crab cavity and the possibility of using various damping schemes to suppress them effectively.

## INTRODUCTION

The finite crossing angle (~ 20 mrad) operation of the CLIC collider needs a crab cavity in the beam delivery system to regain the luminosity loss. A crab cavity is operated in the lowest dipole mode, which has $TM_{110}$-like field patterns, as it has the highest transverse geometric shunt impedance. The operating frequency is chosen as 11.9942 GHz, same as that of the main linac [1]. Hence for 3 TeV CoM beams, crossing at 20 mrad, the cavity will require a kick voltage of 2.4 MV as per Eq. (2) [2]. In addition to providing the necessary kick voltage, means for controlling the wakefields in the crab cavity also should be provided.

Multi-bunch wakefields result in the quality degradation of colliding bunches and the consequent luminosity loss. The longitudinal wakes cause bunch energy gain or loss proportional to the shunt impedance and frequency of the mode exited while the transverse wakes have two effects- (1) impart transverse deflection to the bunches which cause emittance dilution at its minimum and beam break-up instability at its maximum, (2) cause uneven bunch rotation. Either of these effects reduces the luminosity level regained by the crab cavity. One possible way of controlling the wakefields is given by the CLIC baseline design where a heavily damped quadrant structure is used [3]. In this, the beam induced modes are extracted through specially cut slots on the cavity and dumped in external loads. Another method which was implemented for the NLC, uses a moderately damped structure with dipoe mode detuning. In a detuned structure, the cell dimensions are adjusted along the structure such as to spread the dipole frequency, hence preventing their growth downstream [4]. However the latter is successful only in reducing the effect of the first dipole passband. For the CLIC, the bunch spacing is only 0.5 ns (in a train of 312 bunches) hence a heavily damped ($Q_s < 50$) and detuned structure could be the best option for the crab cavity.

## WAKEFIELD SPECTRUM

The wakefield spectrum of a dipole cavity is different from that of an accelerating cavity. For a dipole cavity, the spectrum consists mainly of the fundamental accelerating mode (lower order mode or LOM), degenerate crabbing mode (same order mode or SOM) and higher order modes (monopole and dipole HOMs). In addition, each higher order dipole mode has two polarisations. All these have different frequencies and field configurations which make the damping quite complex. However, only those modes which are synchronous with the beam ($v_p$=3x10$^8$ m/s) and with significant shunt impedance need to be considered.

Microwave studio [5] eigenmode solver has been used to obtain the mode spectrum of the 11.9942 GHz dipole cavity with $2\pi/3$ phase advance/cell, (a = 5 mm, b = 14.083 mm, L = 8.337 mm, t = 2 mm). The synchronous modes up to the 6$^{th}$ dipole band are identified from the calculated dispersion curves for an infinitely periodic structure, shown in Fig. 1.

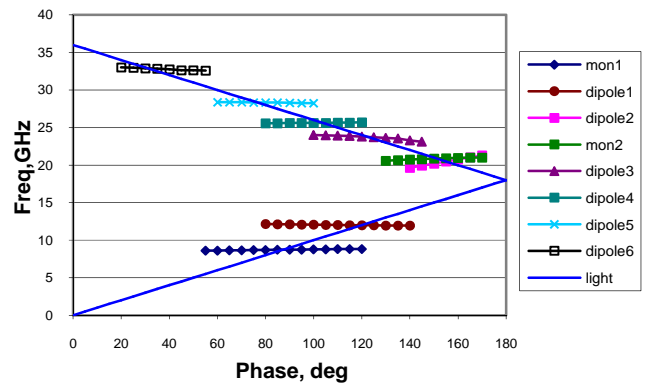

Figure 1: Dispersion curves for the 2 monopole (blue, green) and 6 dipole modes in the CLIC crab cavity

The calculated synchronous R/Q is the largest for the 1$^{st}$ dipole band (hence for the SOM), followed by 3$^{rd}$, 4$^{th}$, 2$^{nd}$ and 6$^{th}$ dipole bands. This suggests very high damping for the SOM and moderate damping for the rest. In addition, the monopole modes (mon 1 and 2 in Fig. 1), offering high shunt impedance also need to be damped.

## DAMPING METHODS

Any beam induced field will sustain in the cavity for a time $\tau = 2Q_L/\omega_s$, where $Q_L$ is the loaded quality factor and

$\omega_s/2\pi$ is the synchronous frequency of the mode [6]. It is the purpose of damping to increase the power loss experienced by that mode in the cavity thereby dissipating its energy either in the cavity walls itself or in some external dielectric load thereby lowering the time $\tau$. A damper in its simpler form consists of coupling slots cut into the cavity in appropriate planes which then extend to an external load through a matched waveguide. The slot and waveguide dimensions are so chosen as not to couple to the operating mode.

Most of the conventional damping methods when used as such are not compatible with a dipole cavity. This is because, one of the two polarisations of the 1$^{st}$ dipole mode, is the operating (crabbing) mode that needs to be maintained while the other is the most problematic SOM that needs to be removed. Hence it is desired to make the cavity or the damping scheme asymmetric thereby isolating the operating mode from the extraction of the SOM. In the following section, we will focus on the asymmetrical choke damper and waveguide damper designs.

*Asymmetrical choke damper*

A basic choke is a radial extension to the basic cavity that allows all the unwanted modes to be coupled to and dumped on the ring shaped load, while reflecting back the operating mode [7]. However such a symmetric structure has the drawback that the SOM is left unaffected as the operating mode. However the LOM is damped to a $Q_{ext} < 100$. Introducing an azimuthal asymmetry by removing a small section of the choke of width $w$ in one of the orthogonal planes, will leave the dipole mode defined by that plane coupled to the load. The asymmetrical choke-mode cavity is shown in Fig. 2.

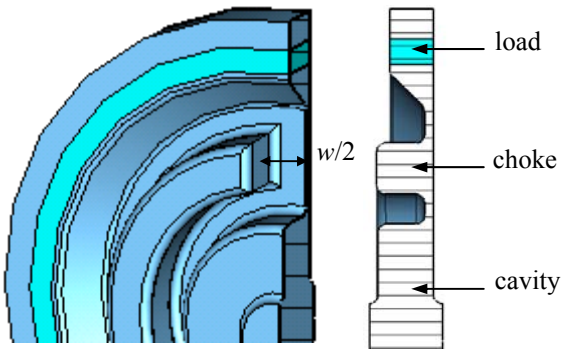

Figure 2: Asymmetric choke-mode cavity

The load material used in simulations has properties $\varepsilon_r = 5$, $\tan\delta = 0.05$. Figure 3 shows the damping performance of the above cavity up to the 6$^{th}$ dipole mode for $w = 10$ mm. The above structure thus damps the SOM to $Q_{ext} \approx 650$ and the other modes in the range similar to a symmetric choke. The level of choke-asymmetry (notch-width) determines the depth of SOM damping. Note that the operating or crab mode is unaffected, however increasing the asymmetry of the structure will increase the losses of the crab mode also.

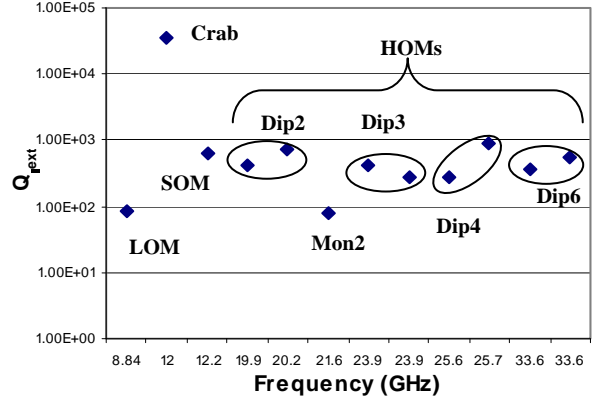

Figure 3: $Q_{ext}$'s of the modes in an asymmetrical choke-damped cavity. (degenerate modes are encircled)

*Waveguide damper*

This type of damper is directly followed from the CLIC baseline design. As far as a crab cavity is concerned, a WG damper has the main advantage of polarizing the cavity due to its asymmetrical nature. It uses a cut-off waveguide (WG) attached to the equator on one or both orthogonal planes of the cell and is terminated by lossy dielectric loads. Hence, this will effectively damp the LOMs, SOM and some of the dipole HOMs, having strong fields near the circumference while preserving the operating mode which has field strength in the orthogonal plane. A 40 mm long rectangular waveguide of suitable cross section (22x8.337 mm$^2$) is chosen giving a TE$_{10}$ cut-off below the LOM frequency. The coupling width $w_1$ was properly chosen to maintain a high $Q_{ext} \approx 1.0\text{x}10^8$ for the operating mode. The LOM and SOM are damped respectively to $Q_{ext}$ ~200, and 100.

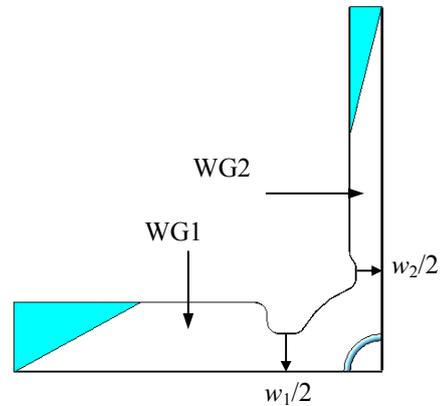

Figure 4: Four-port waveguide damper

However, some TE-like HOMs are moderately damped to ~ 1000. These TE-like modes can be extracted out of the cavity by using a second waveguide (WG2) in the orthogonal plane and which is cut-off to the operating mode. The structure is shown in Fig. 4. The dimensions of WG2 is a_wg = 10 mm with a $TE_{10}$ cut-off above 12 GHz. For $w_1$ = 13 mm and $w_2$ = 9 mm, the Dip2 $Q_{ext}$ is reduced to 170. The waveguide damping performance is shown in Fig. 5.

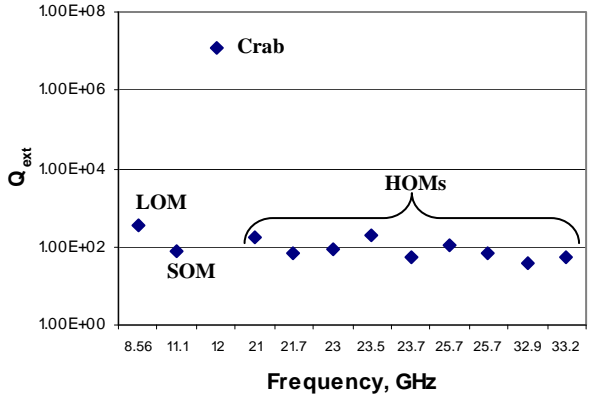

Figure 5: $Q_{ext}$'s of the modes in a waveguide-damped cavity

Above results reveal that the waveguide damper is far superior to the choke damper in terms of damping the SOM and the HOMs. It also provides lesser leakage of the operating mode from the cavity into the damper. But for the LOM, the choke damper performs better because of its structure that is near-symmetric compared to the latter.

## A MULTI-CELL DAMPED CRAB CAVITY

A 7-cell waveguide damped crab cavity is designed with matched input and output couplers as shown in Fig. 6. The 1st and 7th cells have the symmetric waveguide couplers for the operating mode and the rest have the waveguide dampers on them. The damped cell dimensions have been chosen from the periodic cell simulation explained above. The coupler is then optimised for field flatness and phase advance/cell as plotted in Fig. 6 and 7.

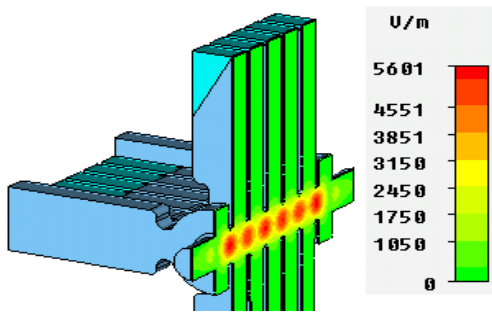

Figure 6: A 7-cell damped travelling wave crab cavity

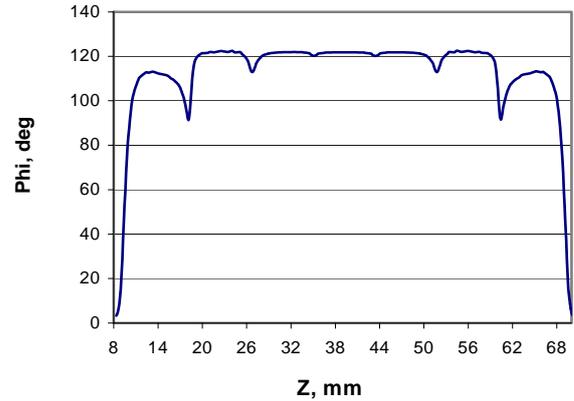

Figure 7: Phase advance/cell along the structure

## CONCLUSION

Wakefield damping in the CLIC crab cavity requires special consideration of the damper design. Asymmetrical dampers are the only solution as they polarise the cavity thereby separating the orthogonal dipole modes. A choke damper is attractive from the fabrication point of view and due to its well-damping of LOM ($Q_{ext}\approx$90). But its moderate SOM damping ($Q_{ext}\approx$700) and low isolation to the operating mode ($Q_{ext}\approx$3.3e4) are undesirable. On the other hand, a waveguide damper has the above Q's as $\approx$350, 80 and 1.2e7 respectively for the LOM, SOM and the operating mode. Hence the latter is preferred for the crab cavity as transverse wakes are more dangerous than longitudinal wakes. Also by using a lossier load material, the desired level of heavy damping can be achieved.

## ACKNOWLEDGEMENT

The work is supported by the Science and Technology Facilities Council (STFC).